\documentclass[12pt]{article}
\usepackage{cite,amssymb,graphicx,graphics}
\begin{document}
\title{\bf Localization of scalar and tensor fields in the standing wave braneworld with increasing warp factor}
\author{{\bf Merab Gogberashvili}\\
Andronikashvili Institute of Physics, \\
6 Tamarashvili St., Tbilisi 0177, Georgia \\
and \\
Javakhishvili State University, \\
3 Chavchavadze Ave., Tbilisi 0128, Georgia\\
{\sl E-mail: gogber@gmail.com} \\\\
{\bf Pavle Midodashvili}\\
Ilia State University, \\ 3/5 Kakutsa Cholokashvili Ave., Tbilisi 0162, Georgia\\
{\sl E-mail: pmidodashvili@yahoo.com} \\\\
{\bf Levan Midodashvili}\\
Gori University, \\ 53 Chavchavadze St., Gori 1400, Georgia \\
{\sl E-mail: levmid@hotmail.com}}
\maketitle
\begin{abstract}
We investigate scalar and tensor fields in the brane model solution for the 5D space-time with standing gravitational waves in the bulk and show that even in the case of increasing warp factor there exist normalizable zero modes localized on the brane.
\end{abstract}

\vskip 0.5cm

\section{Introduction}

The scenario where our world is associated with a brane embedded in a higher dimensional space-time with non-factorizable geometry \cite{Hi,brane} has attracted a lot of interest with the aim of solving several open questions in modern physics. Most of these models were realized as time independent field configurations. However, there have appeared several braneworld models that assumed time-dependent metrics and fields \cite{S}. Here we consider the braneworld scenario with non-stationary metric coefficients recently proposed in \cite{Wave}. The braneworld is generated by 5D standing gravitational waves coupled to a phantom-like scalar field in the bulk.

A key requirement for realizing the braneworld idea is that the various bulk fields be localized on the brane. For reasons of economy and avoidance of charge universality obstruction \cite{DuRuTi} one would like to have a universal gravitational trapping mechanism for all fields. However, there are difficulties to realize such mechanism with exponentially warped space-times. In the existing (1+4)-dimensional models spin $0$ and spin $2$ fields can be localized on the brane with the decreasing warp factor \cite{brane}, spin $1/2$ field can be localized with the increasing factor \cite{BaGa}, and spin $1$ fields are not localized at all \cite{Po}. For the case of (1+5)-dimensions it was found that spin $0$, spin $1$ and spin $2$ fields are localized on the brane with the decreasing warp factor and spin $1/2$ fields again are localized with the increasing factor \cite{Od}. There exist also 6D models with non-exponential warp factors that provide gravitational localization of all kind of bulk fields on the brane \cite{6D}, however, these models require introduction of unnatural sources.

As it was mentioned in \cite{Wave} standing wave braneworld model can provide universal gravitational trapping of zero modes of all kind of fields in case of rapid oscillations of standing gravitational waves in the bulk. To understand the idea we want to remind that standing electromagnetic waves, so-called optical lattices, can provide trapping of various particles by scattering and dipole forces \cite{Opt}. In \cite{quadr} localization was demonstrated through quadruple forces as well. It is known that the motion of test particles in the field of a gravitational wave is similar to the motion of charged particles in the field of an electromagnetic wave \cite{Ba-Gr}. Thus standing gravitational waves could also provide confinement of matter via quadruple forces. As a classical example let us consider the equations of motion of the system of spinless particles in the quadruple approximation \cite{Dix},
\begin{equation} \label{quad}
\frac{Dp^\mu}{ds}= F^\mu_{quad} = -\frac 16 J^{\alpha\beta\gamma\delta}D^\mu R_{\alpha\beta\gamma\delta}~,
\end{equation}
where $p^\mu$ is the total momentum of the matter field and $J^{\alpha\beta\gamma\delta}$ is the quadruple moment of the stress-energy tensor for the matter field. The oscillating metric due to gravitational waves should induce a quadruple moment in the matter fields. If the induced quadruple moment is out of phase with the gravitational wave the system energy increases in comparison with the resonant case and the fields/particles will feel a quadruple force, $F^\mu_{quad}$, which ejects them out of the high curvature region, i.e. it would localize them at the nodes.

In this Latter we show that standing gravitational waves induced by the brane in the bulk really can localize scalar and tensor field zero modes on the brane even for the case of exponentially increasing warp factor.


\section{The model}

In this section we briefly recall the 5D standing wave braneworld model proposed in \cite{Wave}. The braneworld is generated by gravity coupled to a non-self interacting scalar phantom-like field \cite{phantom}, which depends on time and propagates in the bulk. The action of the model has the form:
\begin{equation} \label{action}
S = \int d^5x \sqrt{g} \left[\frac{1}{16 \pi G_5} \left( R - 2\Lambda_{5}\right) - \frac{1}{2}g^{MN}\partial_M \phi \partial_N\phi \right],
\end{equation}
where $G_5$ and $\Lambda _5$ are 5D Newton and cosmological constants respectively.

To avoid the well-known problems with stability, which occur with ghost fields, we can associate the ghost-like field $\phi$ with the geometrical scalar field in a 5D integrable Weyl model \cite{Weyl}. In the Weyl model a massless scalar appears through the definition of the covariant derivative of the metric tensor,
\begin{equation} \label{D}
D_A g_{MN} = g_{MN}\partial_A \phi ~.
\end{equation}
This is a generalization of the Riemannian case where the covariant derivative of the metric is zero. The assumption (\ref{D}) implies that the length of a vector is altered by parallel transport. Weyl's scalar field in (\ref{D}) may imitate a massless scalar field - either an ordinary scalar or ghost-like scalar. The gravitational action for the Weyl 5D integrable model can be written as
\begin{equation}\label{grav-weyl}
S = \int d^5x \sqrt{g} \left[\frac{1}{16 \pi G_5} \left( R - 2\Lambda_{5}\right) - (6-5\xi)g^{MN}\partial_M \phi \partial_N\phi  \right]~,
\end{equation}
where $\xi$ is an arbitrary constant. For example, for $\xi = 11/10$ the action (\ref{grav-weyl}) exactly coincides with (\ref{action}). Thus we can start with a 5D Weyl model and require that we have Riemann geometry on the brane by assuming that the geometrical scalar field $\phi$ is independent of 4D spatial coordinates, $x^i$, and vanishes on the brane. The definition of $\phi$ via (\ref{D}) avoids the usual instability problems of ghost fields since the geometrical fields have specific couplings with matter fields and it is known that the Weyl model is stable for any value of $\xi$.

The Einstein equations for the action (\ref{action}) read:
\begin{equation} \label{field-eqns}
R_{MN} - \frac{1}{2} g_{MN} R =  8\pi G_5 T_{MN} - \Lambda _5 g_{MN}~.
\end{equation}
The metric {\it ansatz} we use is:
\begin{equation} \label{metricA}
ds^2 = e^{2a|r|}\left( dt^2 - e^{u}dx^2 - e^{u}dy^2 - e^{-2u}dz^2
\right) - dr^2~,
\end{equation}
where $a$ is a constant, which corresponds to brane width. The peculiarity of the model (\ref{metricA}) is that the brane, located at $r = 0$, possesses anisotropic oscillations and send a wave into the bulk (as in \cite{gms}), i.e. the brane is warped along the spatial coordinates $x, y, z$ through the factor $e^{u(t,r)}$, which depends on time $t$ and the extra coordinate $r$.

On the other hand, the phantom-like scalar field $\phi (t,r)$ obeys the Klein-Gordon equation on the background space-time given by (\ref{metricA}),
\begin{equation} \label{phi}
\frac{1}{\sqrt{g}}~\partial_M (\sqrt{g} g^{MN}\partial_N \phi)
= e^{-2a|r|}\ddot \phi - \phi'' - 4 a \epsilon(r) \phi' = 0 ~,
\end{equation}
where $\epsilon(r)$ is the sign function and overdots and primes mean derivatives with respect to $t$ and $r$ respectively.

We further rewrite the Einstein equations in the form:
\begin{equation} \label{field-eqns1}
R_{MN} = - \partial_M \sigma\partial_N
\sigma + \frac{2}{3} g_{MN} \Lambda _5~,
\end{equation}
where the gravitational constant has been absorbed in the definition of the scalar field:
\begin{equation}
\label{sigma} \sigma = \sqrt{8 \pi G_5} ~ \phi ~.
\end{equation}
It turns out that the system (\ref{field-eqns1}) is consistent only when the fields $\sigma$ and $u$ are related \cite{Wave}:
\begin{equation} \label{sigma=u}
\sigma (t,r) = \sqrt{\frac{3}{2}}~u(t,r)~.
\end{equation}

The anisotropy of the metric (\ref{metricA}) forces us to define the energy-momentum tensor of the brane with different tensions along different directions:
\begin{eqnarray} \label{tensormixto}
\tau^{M}_{N}= \delta{(r)}~\mbox{\rm
diag}[\lambda_t,\lambda_x,\lambda_y, \lambda_z,0]~, ~~~~~ (\lambda_x
= \lambda_y)~.
\end{eqnarray}
The tensions $\lambda_m$ ($m=t,x,y,z$) in general depend only on time. Taking into account (\ref{tensormixto}) the field equations (\ref{field-eqns1}) change as follows:
\begin{equation} \label{field-eqnsDELTA}
R_{MN} = - \partial_{M}
\sigma\partial_{N} \sigma + \frac{2}{3} g_{MN} \Lambda _5 + 8\pi G_5 \bar{\tau}_{MN}~,
\end{equation}
where the reduced energy-momentum tensor,
\begin{equation}
\bar{\tau}_{MN} = \tau_{MN} - \frac{1}{3}g_{MN}\tau~,
\end{equation}
corresponds to the matter content on the brane and takes the form:
\begin{eqnarray}\label{reduced}
\bar{\tau}_{MN}=\frac{1}{3}\delta{(r)}~\mbox{\rm diag}\left[
2\lambda_t + 2e^{-u}\lambda_x + e^{2u}\lambda_z ,e^{u}\lambda_t + \lambda_x - e^{3u}\lambda_z, \right.\\
\left. e^{u}\lambda_t + \lambda_x - e^{3u}\lambda_z, e^{-2u}\lambda_t - 2e^{-3u}\lambda_x + 2\lambda_z, 0\right]~. \nonumber
\end{eqnarray}

The non-zero components of the Ricci tensor for the metric (\ref{metricA}) read:
\begin{eqnarray} \label{ricci}
R_{tt} &=& e^{2a|r|} \left[-\frac {3}{2} e^{-2a|r|} \dot u ^2 + 4a^2 + 2a\delta(r)\right]~, \nonumber \\
R_{xx}&=&R_{yy}= e^{2a|r|+u}\left[ \frac {1}{2} e^{-2a|r|}\ddot u -
4a^2 - 2a\epsilon(r)u' - 2a\delta(r)
-\frac {1}{2} u'' \right]~, \nonumber \\
R_{zz} &=& e^{2a|r|-2u}\left[ - e^{-2a|r|}\ddot u - 4a^2 + 4a\epsilon(r)u' - 2a\delta(r) + u''\right]~, \\
R_{rr} &=& -\frac 32 u'^2 - 4a^2 - 8a\delta(r)~, \nonumber\\
R_{rt} &=& -\frac 32 \dot uu' ~. \nonumber
\end{eqnarray}
Using the fine tuning:
\begin{equation}\label{L=a}
\Lambda_ 5 = 6 a^2~,
\end{equation}
the system of Einstein equations (\ref{field-eqnsDELTA}) reduces to a single ordinary differential equation,
\begin{equation}\label{eqntou}
e^{-2a|r|}~\ddot u - u'' - 4a\epsilon(r) u' = 0~.
\end{equation}
In addition, for the brane tensions $\lambda_m$ we have \cite{mm}:
\begin{eqnarray}
\lambda_t &=& -\frac{3a}{4\pi G_5} \delta \left(r\right)~, \nonumber \\
\lambda_x &=& \lambda_y = \frac{e^{u\left(t,0\right)}}{8\pi G_5} \left(6a-\frac{1}{2}[u']\right)\delta \left(r\right) ~,\\
\lambda_z &=& \frac{e^{-2u\left(t,0\right)}}{8\pi G_5}\left(6a+[u']\right)\delta \left(r\right)~, \nonumber \\
\lambda_r &=& 0~, \nonumber
\end{eqnarray}
where  $u\left(t,0\right)$ and $[u']$ denote the value of $u(t,r)$ and the jump of its first derivative at $r=0$, respectively.

The standing wave solution to (\ref{eqntou}) can be constructed by implementing the {\it ansatz}~:
\begin{equation} \label{separation}
u(t,r) = C \sin (\omega t) f(r)~,
\end{equation}
where $C$ and $\omega$ are real constants. From (\ref{eqntou}) we get the equation for the radial function:
\begin{equation} \label{f}
f'' + 4 a \epsilon(r) f' + \omega^2 e^{-2 a |r|}f = 0 ~.
\end{equation}
The general solution to this equation has the following form:
\begin{equation} \label{fsol}
f(r) = e^{-2a|r|} \left[A~ J_2\left( \frac{\omega}{|a|} e^{-a|r|}
\right) + B~ Y_2\left( \frac{\omega}{|a|} e^{-a|r|} \right)\right],
\end{equation}
where $A,B$ are arbitrary constants and $J_2$ and $Y_2$ are second-order Bessel functions of the first and second kind, respectively.

As pointed out in \cite{Wave}, along with the solution (\ref{fsol}), the ghost-like field $\sigma (t,r)$ must be unobservable on the brane. Taking into account the relations (\ref{sigma=u}) and (\ref{separation}) we can accomplish this requirement by setting the boundary condition:
\begin{equation}
\left. f (r) \right|_{r=0} = 0~.
\end{equation}
Since $J_2$ and $Y_2$ are oscillatory functions, in the case of increasing (decreasing) warp factor for some fixed values of the constants $A$ and $B$ the function $f(r)$ can have finite (infinite) number of zeros. Thus the above boundary condition can be written in the form which quantizes the oscillation frequency, $\omega$, of the standing wave in terms of the brane width, $1/a$, i.e.
\begin{equation} \label{quantize}
\frac{\omega}{|a|} = X_n~,
\end{equation}
where $X_n$ is the $n^{th}$ zero of the function $f(r)$. Correspondingly, the functions $\sigma (t,r)$ and $u(t,r)$ vanish at the finite (infinite) number of points along the extra dimension. These points - the nodes of standing wave - can be considered as 4D space-time `islands', where the matter particles are assumed to be bound.


\section{Localization of scalar and metric fields}

Since the equations for extra components of tensor and scalar fluctuations are similar for the simplicity let us consider localization problem only for scalar fields, which are defined by the 5D action:
\begin{equation} \label{Sphi}
S_\Phi = - \frac 12 \int \sqrt{g} dx^4dr ~ g^{MN}\partial_M\Phi \partial_N\Phi~.
\label{actionphim0}
\end{equation}
Corresponding Klein-Gordon equation has the form:
\begin{equation}\label{ScalFieldEqn}
\frac{1}{\sqrt g}~\partial_M \left( \sqrt g g^{MN}\partial_N \Phi \right) = 0~.
\end{equation}
Determinant for our {\it ansatz} (\ref{metricA}) is equal to $\sqrt g = e^{4a|r|}$ and (\ref{ScalFieldEqn}) can be written as:
\begin{eqnarray}\label{Equation1}
\left[ \partial_t^2 - e^{- u}( \partial_x^2 + \partial_y^2) - e^{2u} \partial_z^2\right]\Phi = e^{2a|r|}\left( e^{4a|r|} \Phi' \right)',
\end{eqnarray}
where primes denote derivatives with respect to $r$. Let us look for the solution in the form:
\begin{equation}\label{Solution1}
\Phi \left( {t,x,y,z,r} \right) = \Psi (t,r)\chi (x,y)\xi (z)~,
\end{equation}
which transforms (\ref{Equation1}) into the system of the equations,
\begin{eqnarray} \label{system}
\left(\partial _x^2 + \partial _y^2 \right)\chi + \left( p_x^2+p_y^2\right)\chi &=& 0 ~, \nonumber \\
\partial_z^2\xi + p_z^2\xi &=& 0 ~, \\
\partial_t^2\Psi + \left[ (p_x^2 + p_y^2)e^{-u} + p_z^2 e^{2u} \right]\Psi &=& e^{2a|r|}\left( e^{4a|r|}\Psi' \right)'. \nonumber
\end{eqnarray}
In the case $u\approx 0$ the parameters $p_x, p_y$ and $p_z$ can be regarded as momentum components along the brane. Their exact physical meaning is not clear in general case since (\ref{system}) are equations with variable coefficients.

To solve the system (\ref{system}) we assume that the constant $A$ in the solution (\ref{fsol}) is zero and the oscillation function of gravitational waves has the form:
\begin{equation} \label{u}
u \left( t,r \right) = B ~ \sin \left( \omega t \right) e^{-2a|r|} Y_2\left(\frac {\omega}{|a|}~ e^{-a |r| }\right)~,
\end{equation}
where $B$ is a constant and $Y_2$ is the second-order Bessel function of the second kind. Physical solutions described by Bessel functions usually contain the first kind functions, $J_n$, because of their regularity at the origin. However, in our case we can use even Bessel function of the second kind, $Y_n$, since the argument in (\ref{u}), $e^{-a |r|} \omega / |a|$, is always positive and becomes zero only for the case of increasing warp factor ($a > 0$) at $r \to \pm \infty$.

To show explicitly localization by standing waves we consider the extreme case with the increasing warp factor $a>0$. The function (\ref{u}) is zero at the position of the brane, $r = 0$, due to fine tuning (\ref{quantize}), where $X_n$ now is one of the zeros of $Y_2$.

For 4D massless scalar modes on the brane we have the following dispersion relation:
\begin{equation}
E^2 = p_x^2 + p_y^2 + p_z^2~.
\end{equation}
When the frequency of standing gravitational waves is much larger than frequencies associated with the energies of the particles on the brane,
\begin{equation}
\omega \gg E~,
\end{equation}
we can perform time averaging of oscillating exponents in the equation for scalar field (\ref{system}).
Using the expressions:
\begin{eqnarray} \label{e-br}
\frac{\omega }{2\pi} \int\limits_0^{2\pi/\omega} \left[ \sin (\omega t) \right]^m dt &=& \left \{
\begin{array} {lr}
0 & (m = 2n + 1)\\\\
2^{-2n}\left( 2n \right)! \left( n! \right)^{-2} & (m = 2n)
\end{array}\right. \nonumber \\
e^{bu} &=& \sum \limits_{n = 0}^{+\infty } \frac{\left( bu \right)^n}{n!}~,
\end{eqnarray}
where $b$ is an integer, for the time averages of oscillating exponents (\ref{e-br}) we get:
\begin{equation}
\left\langle e^{bu} \right\rangle =  1 + \sum\limits_{n = 1}^{ + \infty } \frac{(bB)^{2n}e^{- 4an|r|} }{2^{2n}\left( n! \right)^2} \left[ Y_2\left( \frac{\omega }{a}e^{ - a|r|} \right) \right]^{2n} = I_0(z)~,
\end{equation}
where $I_0$ is the modified Bessel function of the zero order with the argument:
\begin{displaymath}
z =|b B| \left[ \frac{2a}{\omega} e^{-a|r|}Y_1 \left(\frac{\omega}{a}e^{-a|r|}\right)- e^{-2a|r|} Y_0\left(\frac{\omega}{a}e^{-a|r|}\right) \right] .
\end{displaymath}

Now separating variables,
\begin{equation}
\Psi ( t,r) = e^{iEt} \rho (r)~,
\end{equation}
for the extra factor $\rho (r)$ of the wave function we obtain:
\begin{equation}\label{rho}
\left( e^{4a|r|}\rho' \right)' - e^{2a|r|}F(r) \rho  = 0~,
\end{equation}
where
\begin{equation}
F(r) = \left(\left\langle e^{-u} \right\rangle -1\right)\left(
p_x^2 + p_y^2 \right) + \left(\left\langle e^{2u} \right\rangle
-1\right)p_z^2 ~.
\end{equation}

It is more convenient to put Eq. (\ref{rho}) into the form of an analogue non-relativistic quantum mechanical problem by making the change:
\begin{equation} \label{ro}
\rho (r) = e^{-2a|r|} \psi ( r )~.
\end{equation}
For $\psi (r)$ we find:
\begin{equation}\label{psi}
\psi'' - U(r) \psi = 0~,
\end{equation}
where the function,
\begin{equation}\label{U}
U(r)= 4a\delta (r) + 4a^2 + e^{-2a|r|}F(r) ~,
\end{equation}
is the analog non-relativistic potential. FIG. 1 shows the behaviour of $ U (r)$ in the case of first zero of the Bessel function $Y_2$, i.e. when
\begin{equation}\label{FirstZerosY}
\frac{\omega}{a} \approx 3.38~.
\end{equation}


\begin{figure}[ht]
\begin{center}
\includegraphics[width=0.5\textwidth]{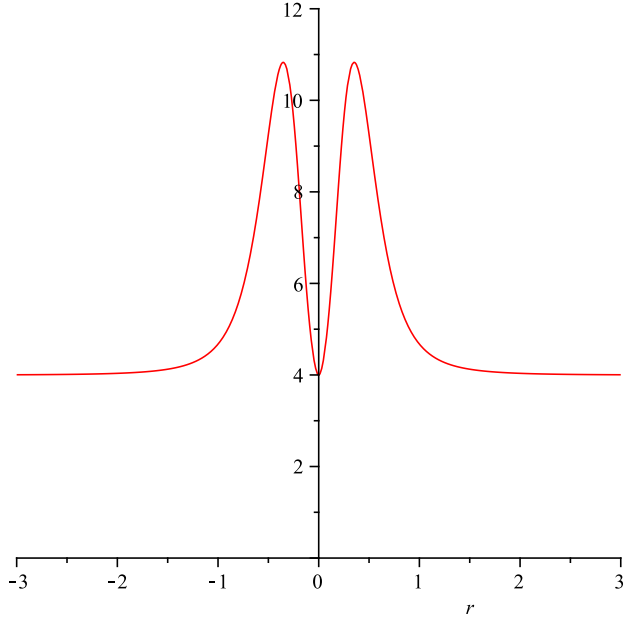}
\caption{The total gravitational potential for scalar particles in the bulk, (\ref{U}).}
\end{center}
\end{figure}


To study general behaviour of extra part of the scalar zero mode wave function we explore Eq. (\ref{psi}) in two limiting regions, far from and close to the brane.

Far from the brane, $r \to \pm \infty$, Eq. (\ref{psi}) has the following asymptotic form:
\begin{equation}\label{infinity}
\psi'' - 4a^2 \psi = 0~,
\end{equation}
with the solutions $\psi \sim e^{\pm 2a|r|}$. Taking into account the normalization condition for the wave function and  $a>0$, we choose $\psi \sim e^{- 2a|r|}$. So, far from the brane for the extra factor of the wave function (\ref{ro}) we obtain:
\begin{equation}\label{rho-infinity}
\rho (r)_{r \to \pm \infty } \sim e^{ - 4a|r|}~.
\end{equation}

In the second limiting region, $r \to \pm 0 $, i.e. near the brane, Eq. (\ref{psi}) can be approximated as:
\begin{equation}\label{zero}
\psi'' - \left[ 4a \delta (r) + 4a^2 + c^2r^2 \right]\psi = 0~,
\end{equation}
where we have introduced the constant:
\begin{equation}
c^2 = \frac{B ^2\omega ^2}{16}\left( p_x^2 + p_y^2 + 4p_z^2 \right) \left[ Y_1 \left( \frac{\omega }{a} \right) - Y_3 \left( \frac{\omega }{a} \right) \right]^2.
\end{equation}
The solution of (\ref{zero}) is:
\begin{eqnarray}
\psi \left(r\right) = \frac{D}{\sqrt{|r|}}\left[ W\left( - \frac{a^2}{c},\frac 14,c|r|^2 \right) + \left(\frac{2\sqrt{\pi}}{\Gamma \left( \frac 14 + \frac{a^2}{c} \right)} + \right.\right. \nonumber \\
\left. \left. + \frac{2\sqrt{\pi}}{\Gamma \left( \frac 34 + \frac{a^2}{c} \right)}\frac{a}{\sqrt{c}}\right)
M \left( - \frac{a^2}{c},\frac 14,c|r|^2 \right)\right]~,
\end{eqnarray}
where $D$ is a constant, $\Gamma$ is the gamma function and $W$ and $M$ denote Whittaker functions. Imposing boundary condition at the origin,
\begin{equation}
\left. \rho' (r)\right|_{r=0} = 0~,
\end{equation}
close to the brane extra part of scalar wave function will behave as (see FIG. 2):
\begin{equation} \label{rho-zero}
\rho (r)|_{r \to \pm 0 } = \frac{D\sqrt{\pi}c^{1/4}}{\Gamma \left( \frac 34 + \frac{a^2}{c} \right)}\left[1 + \frac{c^2r^4}{12} - \frac{ac^2|r|^5}{15} + O\left( r^6\right)\right]~.
\end{equation}

So the extra part of scalar field zero-mode wave function $\rho(r)$ has maximum at $r=0$, falls off from the brane and turns into the asymptotic form (\ref{rho-infinity}) at the infinity.


\begin{figure}[ht]
\begin{center}
\includegraphics[width=0.5\textwidth]{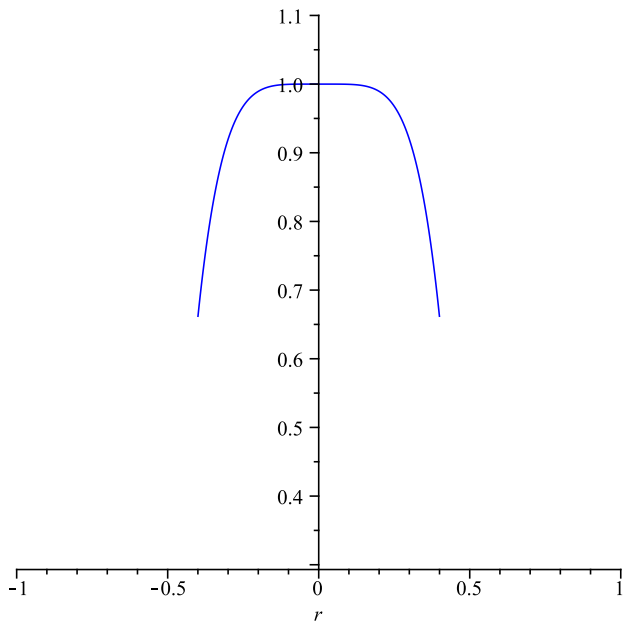}
\caption{The extra part of scalar zero mode wave function close to the brane, (\ref{rho-zero}).}
\end{center}
\end{figure}


In the action of scalar fields (\ref{Sphi}) the determinant, $\sqrt g = e^{4a|r|}$, and the metric tensor with upper indices, $g^{MN}$, give the total exponential factor $e^{2a|r|}$, which obviously increases for $a > 0$. This is the reason why in the original brane models \cite{brane} the scalar field zero modes, with the constant $r$-depended extra part of the wave function, can be localized on the brane only in the case of decreasing warp factor (i.e. $a<0$). In our model the extra part of wave function (\ref{ro}) is not a constant, moreover, for $a>0$ it contains the exponentially decreasing factor $e^{-2a|r|}$.


\section{Conclusion}

In this Latter we have investigated localization problem of scalar and tensor fields within the 5D standing wave braneworld proposed in \cite{Wave}. Since equations for the extra part of wave functions of scalar and tensor fields are similar for simplicity we have explicitly considered in details only the case of scalar fields. We have shown existence of normalizable zero modes on the brane even for the case of increasing warp factor. Further investigations towards the localization of spinor and vector fields in this model are in progress.

\medskip


\noindent {\bf Acknowledgments:} The research was supported by the grant of Shota Rustaveli National Science Foundation $\#{\rm GNSF/ST}09\_798\_4-100$.


\end{document}